# Attila Lajos Makai

# Startup ecosystem rankings

ATTILA LAJOS MAKAI
PhD Student
Széchenyi István University,
Hungary
Email: makai.attila.lajos@sze.hu

The number, importance, and popularity of rankings measuring innovation performance and the strength and resources of ecosystems that provide its spatial framework are on an increasing trend globally. In addition to influencing the specific decisions taken by economic actors, these rankings significantly impact the development of innovation-related policies at regional, national, and international levels. The importance of startup ecosystems is proven by the growing scientific interest, which is demonstrated by the increasing number of related scientific articles. The concept of the startup ecosystem is a relatively new category, the application of which in everyday and scientific life has been gaining ground since the end of the 2000s. In parallel, of course, the demand for measurability and comparability has emerged among decision-makers and scholars. This demand is met by startup ecosystem rankings, which now measure and rank the performance of individual ecosystems on a continental and global scale. However, while the number of scientific publications examining rankings related to higher education, economic performance, or even innovation, can be measured in the order of thousands, scientific research has so far rarely or tangentially addressed the rankings of startup ecosystems. This study and the related research intend to fill this gap by presenting and analysing the characteristics of global rankings and identifying possible future research directions.

In the literature on innovation systems and their spatial extent, the concept of ecosystems (both as a unit of analysis and as a distinct, definable spatial category) is gaining popularity. The term ecosystem was first introduced to social science by *James F. Moore* in a *Harvard Business Review* article in the 1990s. He argues that businesses do not develop in a 'vacuum' environment but rather in the process of interaction between suppliers, customers, and financiers, which is very similar to ecosystems found in nature and thus he borrows the name from ecology, calling the system an economic ecosystem (*Moore* [1993]). The economic ecosystem as a concept has relatively quickly become part of the academic (and subsequently policy) discourse, although its meaning is constantly expanding and evolving, notably due to the related theoretical and modelling scientific advances. A range of theoretical models of entrepreneurial ecosystems have been developed in recent years and a lifecycle typology of ecosystems has been introduced (*Brown–Mason* [2017]).

Venture capital flowing into startup ecosystems seeking profitability, founders facing extreme uncertainties and risks, and public policy-makers allocating significant public funding to the sector naturally want to obtain as much information as possible about the resources available in a region, the amount of capital available for investment, and the survival and potential development paths of startup companies in the area. This need for information is served, on the one hand, by databases that focus specifically on this sector (e.g. Crunchbase, Dealroom, PitchBook, etc.) and on the other hand, by rankings measuring the individual ecosystems. The global startup ecosystem rankings, published in 2016–2017, have received considerable attention and coverage in recent years, both in the press and social media platforms. The released findings and the rankings published every year generate much interest in the startup community and among investors. Also, government announcements in the international media arena are often used to justify a country's success in innovation policy by moving up on the startup ecosystem rankings. In contrast to the interest of the economic, communication, and political sectors, the scientific community has not yet devoted comparable attention to the topic. There are only three publications on Web of Science (WoS) that include the term in their title, keywords, or abstract,[1] but even these are deficient regarding the structure and methodology of the global rankings. This does not constitute scientific proof of the lack of literature; nevertheless, it is certainly indicative. The main objective of this study is to identify the existing global rankings with a scientific claim, describe the ranking methodology, and compare the rankings with each other. It also aims to outline a research agenda (similar to publications on higher education, economic, political, and other rankings) to explore the characteristics and impacts of these rankings in more depth. Based on the above, the research questions are as follows:

*1*. What are the global rankings of startup ecosystems?

*2*. What is the methodology used by each ranking, exactly what is measured and what is not measured, and what variables are used to compile the ranking?

*3*. When comparing the rankings, can a correlation be detected between them?

---

[1] WoS search date: 3 October 2021. Search terms: AB = startup ecosystem rank, AB = startup ecosystem ranking, TS = startup ecosystem rank, TS = startup ecosystem ranking (AB: abstract; TS: topic).

# 1. Literature review and research methodology

## 1.1. Ecosystems

There is a clear consensus in the related studies and literature that innovation ecosystems are developed in urban areas. As the network hub of the innovation ecosystem, the city is where the resources, capacities, explicit and tacit knowledge that fundamentally determine the nature and potential performance of the ecosystem are concentrated (*Clark* [2020]). Startup enterprises and specialised communities organised around them constitute an integral part of these ecosystems. There is some conceptual confusion as some publications refer to innovation ecosystems, others to entrepreneurial ecosystems and others to startup ecosystems, depending on the focus and subject of the research. Naturally, these concepts cannot be sharply distinguished from each other as they mostly describe and explain phenomena in the same geographical area, using the same spatial resources. In the case of startup ecosystems, the processes, network connections, resources, and various capacities associated with startup companies are obviously part of the ecosystem concept. Some help in clarifying this concept is provided by systematic literature reviews on the subject. Based on the most recent literature review on this topic (which analysed 63 publications specifically related to the startup ecosystem), the startup ecosystem refers to 'a limited region within 30 miles (or one-hour travel) range, formed by people, their startups, and various types of supporting organisations, interacting as a complex system to create new startup companies and evolve the existing ones' (*Tripathi et al.* [2019] p. 2). A prerequisite for the emergence and development of startup ecosystems is the combined availability of resources, capacities, and capital elements that enable startup founders to create a product and business model with the required growth potential and a team to achieve it (*Feld* [2012] pp. 147–156), and to establish a community that supports the encounter and cooperation between founders and investors (*Cohan* [2018] pp. 83–85).

Entrepreneurial ecosystems have received considerable academic attention from scientific, economic, and policy system actors in recent years. The widely adopted concept has different aspects emphasised in each research; however, a common point of agreement is that the entrepreneurial ecosystem is understood as a community of mutually interested and interdependent parties with specific resources and a supportive environment for the creation of new enterprises and the development of existing ones within a defined geographical area (*Acs–Autio–Szerb* [2014], *Adler et al.* [2019], *Alvedalen–Boschma* [2017], *Audretsch–Belitski* [2017], *Audretsch–Link* [2019], *Auerswald–Dani* [2017], *Autio et al.* [2018], *Budden–Murray–Turskaya* [2019]) .

If we want to explore the rankings associated with the concept, it is also necessary to analyse its individual components. In this way, we can answer the very important question of whether the variables and indicators used in the rankings cover all the concept components in their entirety or whether they only examine subelements and rank ecosystems on the basis of these. When searching for the components of the concept of the entrepreneurial ecosystem, it is advisable to choose the literature research method (*Grant–Booth* [2009]) that provides the most substantive information, with a minimum narrative element, a broad outlook, and a summary synthesis. This criterion is best met by the 'state-of-the-art' or 'systematic search and review' methodology. There is no need to prepare this as the most recent literature review on entrepreneurial ecosystems has been conducted in this way (*Cao–Shi* [2021]), which, based on the corpus analysed, divides ecosystem components into three groups according to the approaches that explore them (it would be beyond the scope of this study to present all the related literature, so in the following only the relevant findings of the referenced research are presented).

The approach that emphasises the ecosystem as a structural system of interactions (*Cao–Shi* [2021] pp. 80–81) and the importance of interactions focuses on the actors involved and their loosely interlinked network rather than on the various resources and their quantitative and qualitative criteria. Accordingly, the related research focuses on interaction content, interaction patterns, and the resulting components and their networks. The most important resource of the interaction approach is the knowledge held by the different actors, which generate and shape innovation within the ecosystem due to their mutual interaction. The quality of the ecosystem will therefore be defined in the knowledge base available in it as well as in the formal and informal interactions that help to share and transfer it.

The resource logic approach (*Cao–Shi* [2021] pp. 81–82) interprets ecosystems as resource allocation systems whose features are strongly influenced by the characteristics of the institutions and organisations holding resources. Each actor is defined as a coordinator influencing the resource flow, combining and exploiting different types of resources (capital, knowledge, infrastructure, labour, etc.) in the context of the entrepreneurial discovery process. The ability to recycle resources, a quality criterion for each ecosystem, is an important element in the concept. Entrepreneurship failure in this approach is not an adverse event as it does not lead to a loss of resources but merely to their reallocation. Similarly, entrepreneurship success (e.g. successful exit) is not seen as the exit of a particular entrepreneur from the ecosystem but rather as emergence in new roles (mentor, angel, investor). On this basis, the dynamics of resource recycling is an area that has been researched extensively and in detail in this approach. The ability of enterprises to mobilise and combine individual ecosystem resources is also an important aspect.

The next approach to interpreting the functioning of individual ecosystems examines and highlights the governance model (*Cao–Shi* [2021] pp. 83–85). The focus here is on governance structures and their effects which have a significant impact on the quality of the ecosystem. The governance logic to align the objectives and activities of different actors (firms, entrepreneurs, investors, universities, governmental bodies) and the manifestation of multi-polar coordination to facilitate stakeholder cooperation within an ecosystem is a very popular research orientation for literature on this subject.

One of the objectives of this study is to review (adopting the above typology) the variables underlying the rankings measuring/evaluating the performance of startup ecosystems, which element they are related to, which component they measure. In this way, the rankings can also be assessed in terms of whether and to what extent they cover the relevant components of the ecosystem.

## 1.2. Rankings and their comparison

Nowadays, indices and rankings measuring different types of performance are present in all social subsystems and their importance is growing among economic, political, and scientific actors and in everyday life (*Muller* [2019]). They significantly influence decision-making processes; their impact on choices and preferences has been proven in related studies (*Erdi* [2020]). They affect personal/institutional/state reputation (*Origgi* [2018]) and play a role in the political decision-making process (*Cooley–Snyder* [2015], *Malito–Umbach–Bhuta* [2017]). The psychological effects of rankings also relate to areas of high scientific interest (*Merry* [2016]).

Commonly, the same phenomenon/area is assessed by several rankings. In this case, how to compare the different rankings is an important scientific problem. In this respect, higher education rankings are probably the most researched field (*Hazelkorn* [2015]) where well-developed and validated methodologies are available for comparing global rankings (*Hertig* [2016]). The analyses relating to these rankings also provide a useful methodology for studying the rankings' validity, consistency, and reliability (*Shin–Toutkoushian–Teichler* [2011] pp. 63–104). *Kendall*'s rank correlation and *Spearman*'s footrule provide adequate methods for comparing ranked lists in statistics (*Langville–Meyer* [2012] pp. 201–210). The comparison is also supported by the definition of overlaps between rankings (*Aguillo et al.* [2010]) and determining an '*M*-value' that can handle non-overlapping elements (*Bar-Ilan–Levene–Lin* [2007], *Fagin–Kumar–Sivakumar* [2003]).

### 1.3. Research methodology

The main element of the research is to explore existing global startup ecosystem rankings. Given that currently no analysis of the startups' ecosystem ranking is available in the academic literature, it is, therefore, a priority to identify these rankings. Some aspects of certain startup ecosystem rankings have already been examined in a previous related study (*Makai–Vasa* [2020]), however, that is not sufficient for a systematic analysis. Google's search engine is a good starting point as from 2012 the 'Knowledge Graph'[2] can be used to retrieve accurate data when searching. It is also necessary to review the most popular startup-related websites and look for links to ecosystem rankings. Nevertheless, these methods are only suitable for obtaining sporadic information. Systematic analysis is aided by the examination of the linkage of relevant keywords to internet pages and the number of hits on these pages, using the tools of the various search engine optimization (SEO) service sites,[3] which allow the comparison of websites and keywords, the search for keywords and their links to internet pages, and the analysis of the number of views on these pages. In this research, the digital performance of the different rankings is compared using Amazon's Alexa service.

Following the scientific identification of global startups ecosystem rankings, analysis of the databases of each ranking is required. The first step is to identify the methodology behind the rankings, followed by exploring the variables. In the second step, the individual rankings are compared, focusing on the overlaps between them and on the Kendall's and Spearman's correlations. The present comparison does not include the previously referred footrule and *M*-value calculations, which will be examined along with the time series analyses later in the research process.

The methodology will answer all three research questions and outline an adequate agenda for future research directions to better understand ecosystem rankings. For reasons of scope, this study only examines the 2020 ranking data and the comparative work will focus on the basic context and correlations, with deeper statistical insights to be explored in future studies.

## 2. Identification of global startup ecosystem rankings

When identifying the existing global startup ecosystem rankings, the research is based on the assumptions that they (like other rankings) should be public, electron-ically accessible, and available to the target groups of the ranking. If all three assumptions are adopted, the inevitable consequence is that these rankings are also searchable on the World Wide Web by keywords. This will make them easily identifiable using the appropriate keywords and search tools. There are applications on the World Wide Web that, once a keyword has been entered, one can identify the keywords most searched by users for that keyword, based on searches and traffic data, and thus the search can be extended to include them. Such an application is KWFinder[4], which provides the associated keywords, presented in Table 1, for the *startup-ecosystem-rank* word combination.

---

[2] https://developers.google.com/knowledge-graph
[3] Examples: alexa.com, semrush.com

Table 1

*Relevant keywords for startup ecosystem rankings*

| Keyword | Average search volume (last 12 months) | Average search volume (last 6 months) | Average search volume (last 3 months) | Location |
|---|---|---|---|---|
| startup ecosystem ranking 2020 | 280 | 140 | 130 | Anywhere |
| state startup ranking 2020 | 220 | 130 | 140 | Anywhere |
| global startup ecosystem report 2020 | 150 | 120 | 100 | Anywhere |
| global startup ecosystem ranking 2020 | 90 | 110 | 130 | Anywhere |
| global startup ecosystem ranking | 120 | 120 | 170 | Anywhere |
| global startup ecosystem index | 140 | 270 | 520 | Anywhere |
| startup city ranking | 50 | 40 | 40 | Anywhere |
| startupblink startup ecosystem ranking | 30 | 30 | 20 | Anywhere |
| startup ecosystem ranking 2021 | 40 | 90 | 190 | Anywhere |
| startupblink report | 20 | 40 | 60 | Anywhere |
| global startup index | 30 | 40 | 80 | Anywhere |
| startup ecosystem index | 20 | 30 | 60 | Anywhere |
| world startup ranking | 20 | 20 | 30 | Anywhere |
| startup ecosystem ranking 2019 | 10 | 10 | 10 | Anywhere |
| startup genome ranking | 10 | 10 | 10 | Anywhere |
| global startup ecosystem ranking 2019 | 10 | 10 | 10 | Anywhere |
| global startup ranking | 10 | 10 | 20 | Anywhere |

*Source*: Own elaboration using KWFinder.

The query shows that the most common keywords related to this topic are: global, startup, ecosystem, rank(ing), and index. Based on this query, the results already include two ranking names (StartupBlink, Startup Genome) as keywords.

[4] https://kwfinder.com

However, there may still be global rankings that are not yet widespread enough to be interpreted as a relevant keyword by the search algorithm. This requires an analysis of the most common keyword combinations that lead to pages based on searches and communication on the internet. Amazon's Alexa SEO application can perform this search, among others. Since three of the most common keywords (global, startup, and ecosystem) have an 'AND' type of correspondence and two (rank[ing] and index) have an 'OR' type of correlation, it is recommended to run the compound search in two versions as follows:[5]

– global startup ecosystem index,
– global startup ecosystem rank(ing).

The ratio of keyword combinations (as batched brands) displayed in the SERP (search engine results page) is presented in Table 2.

Table 2

*Share of consumer access for specific keywords on the World Wide Web (date of query: 7 October 2021)*

| Global startup ecosystem index | | | Global startup ecosystem rank(ing) | | |
|---|---|---|---|---|---|
| Site | Share of voice (%) | Global rank of site | Site | Share of voice (%) | Global rank of site |
| startupblink.com | 44.57 | 95,896. | startupgenome.com | 39.37 | 188,341. |
| crunchbase.com | 14.42 | 2,703. | crunchbase.com | 13.19 | 2,703. |
| startupgenome.com | 11.02 | 188,341. | startupblink.com | 12.73 | 95,896. |
| adda247.com | 6.51 | 3,763. | thehindubusinessline.com | 5.14 | 6,029. |
| thehindubusinessline.com | 3.57 | 6,029. | adda247.com | 4.56 | 3,763. |
| medium.com | 2.95 | 145. | google.com | 4.50 | 1. |
| affairscloud.com | 1.74 | 22,819. | queensu.ca | 3.56 | 14,882. |
| youtube.com | 1.71 | 2. | usc.edu | 2.59 | 3,434. |
| edristi.in | 1.39 | 194,413. | insightsunboxed.com | 2.36 | 1,999,020. |
| business-standard.com | 1.22 | 2,071. | affairscloud.com | 1.77 | 22,819. |
| | | | financialexpress.com | 1.71 | 2,305. |

*Note.* The table shows voice shares of at least 1%.
*Source*: Own elaboration using Amazon's Alexa application.

Table 2 shows only the share of at least 1% hit but this is sufficient for the analysis. Given that only startupblink.com and startupgenome.com can be defined as

---

[5] As the artificial intelligence-based algorithm within Amazon's Alexa can handle all possible sequences of keywords at the same time, there is no need to run additional searches.

78                                                                                            ATTILA LAJOS MAKAIrankings in the hit list, it can be concluded that there are currently two global startup ecosystem rankings that meet the criteria outlined previously (i.e. public, electronically accessible, and searchable by target groups). These are the *StartupBlink* and *Startup Genome* rankings, the characteristics of which are described in sub-sections 3.1. and 3.2.[6]

## 3. Characteristics of global startup ecosystem rankings

### 3.1. Startup Genome ranking

Startup Genome, represented in San Francisco, Berlin, and Delhi (and therefore, has a global coverage), has been publishing global, regional, and sectoral rankings and reports on startup ecosystems since its launch in 2012.[7] Its integration in global startup networks is well illustrated by its cooperation with prominent players in the field (Crunchbase, Dealroom, Global Entrepreneurship Network). The annual report is released online as a web publication and includes the current global ranking, regional and industry summaries, and specific rankings for individual startup industries. Main features of the ranking:

– *Availability*: The annual report is accessible online free of charge after registration, whereas access to the database and variables on which the rankings are based is subject to a fee.
– *Ranking*: This global ranking lists 140 startup ecosystems (top 40 and emerging top 100 ecosystems).
– *Methodology*: The ranking methodology is fully transparent, with a separate methodology chapter in the annual report (*Startup Genome* [2020] pp. 177–186) that includes all the associated weights used for the ranking.
– *Variables*: The applied variables (see Figure 1) are typically 'resource' variables; only two variables measure interactions within the ecosystem, none measures governance elements, so the latter elements only have an indirect impact on the ranking (relation to the performance indicator). The indicators associated with performance and

---

[6] This does not mean that there cannot be rankings for internal use by policymakers or economic decision-makers, though as these are not available and do not have an impact on a wide range of target groups, they fall outside the scope of this study.
[7] https://startupgenome.com/all-reports

HUNGARIAN STATISTICAL REVIEW, VOLUME 4, NUMBER 2, pp. 70–94. DOI: 10.35618/hsr2021.02.en070

investment are based on relevant data on startups and operating investors from the Crunchbase and Dealroom databases, which can be easily verified and traced. However, they have the disadvantage that these databases are not official statistics; the data supply is completely voluntary, that is, only what is uploaded by the users (startups and investors) is visible. Thus, the ranking of a startup ecosystem depends to a large extent on the willingness of investors and startup companies in that ecosystem to voluntarily provide data to the databases. Exit events are an exception as they are easily quantifiable from other sources due to their nature and publicity. An example of this is the initial public offering (IPO), that is, the admission of companies to the stock exchange. Naturally, it also implies that the data used for the ranking and extracted from the above-mentioned databases are not representative. A similar phenomenon can be observed for the indicators measuring interconnection where events uploaded only to specific databases are considered. Overall, the indicators used, apart from the indicators on exits and those derived from official statistics (e.g. number of STEM [science, technology, engineering, and mathematics] students studying in the field, the average salary of IT [information technology] developers, number of registered intellectual properties [Ip-s], average *H*-index of researchers, etc.), cannot be considered as complete or representative and therefore the composite variables derived from them cannot be regarded as such either. The resulting list will necessarily be a hybrid ranking based on the startup ecosystem's real performance and digital visibility.

80  ATTILA LAJOS MAKAI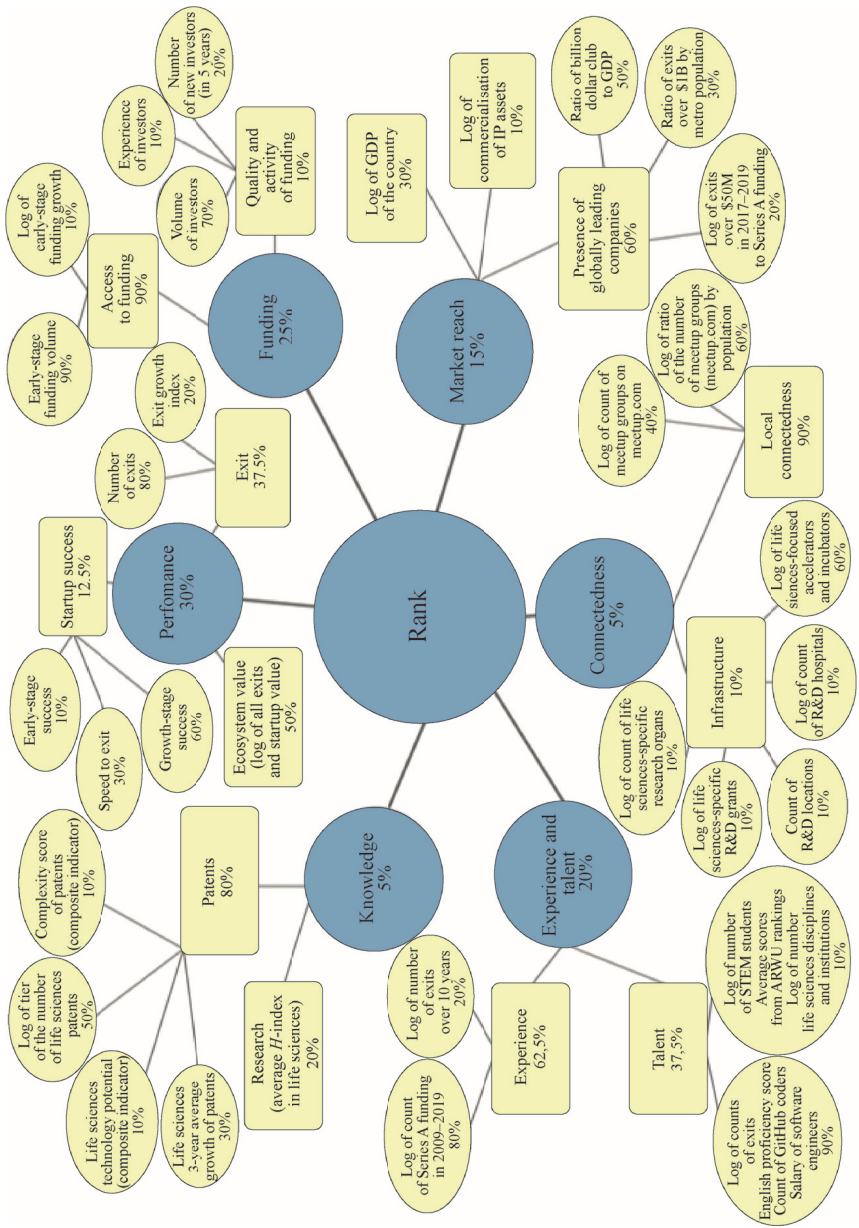

*Figure 1. Startup Genome ranking variables and their weighting*

*Note.* GDP: gross domestic product; prof: proficiency; IP: intellectual property; ARWU: Academic Ranking of World Universities; R&D: research and development.

HUNGARIAN STATISTICAL REVIEW, VOLUME 4, NUMBER 2, pp. 70–94. DOI: 10.35618/hsr2021.02.en070

## 3.2. StartupBlink ranking

The Zurich-based StartupBlink has been publishing an annual report evaluating, presenting, and ranking global startup ecosystems since 2017. A global network of partners (*StartupBlink* [2020] p. 7) and information databases (Crunchbase, SEMrush, Meetup, Coworker.com) closely linked to the startup theme provide the data for the rankings and reports (*StartupBlink* [2020] p. 5). Another specific feature of StartupBlink is that, in addition to publishing reports, it also maintains its web platform where data is available not only on ecosystems but also on individual startups, accelerators, and coworking offices.[8] The StartupBlink databases use the databases of crunchbase.com and coworker.com, while it is also possible to upload the data of a particular startup or incubator directly here. It is equally noteworthy that the Google Maps tool used by the platform provides a direct map display of the information searched for. Main features of the ranking:

– *Availability*: The annual report is accessible online free of charge after registration, whereas access to the database and variables on which the rankings are based is subject to a fee.
– *Ranking*: The StartupBlink Global Ranking lists 1,000 startup ecosystems (and publishes aggregated country rankings).
– *Methodology*: The ranking methodology is algorithm-based and partially transparent, with a separate methodology chapter in the annual report (*StartupBlink* [2020] pp. 10–15) containing the indicators used for the ranking, though the algorithms used are not presented, nor are the weights assigned to each variable. (StartupBlink will not provide any information on the applied ranking methodology, even for the fee-based database.) For this reason, without knowledge of the methodology, the results of the ranking cannot be verified, nor can the characteristics of the algorithms be revealed.
– *Variables*: The variables underlying the ranking are summarised in Figure 2. The variables used are also 'resource' variables; in this case, two of the variables for interactions can be identified. Variables related to the measurement of governance do not appear in this ranking either. For the StartupBlink indicators, concerns about the data derived from voluntary databases as already presented for Startup Genome, can also be mentioned. It should also be pointed out that particular emphasis is given to indicators derived from other indices for the variables relating to the business environment. This may distort the ranking in two ways.[9] On the one hand, the methodological shortcomings of the index used are incorporated into the ranking. This is a special concern for a compromised index such as the World Bank's doing business index (DBI). On the other hand, the inclusion of certain political content indices in the evaluation can also have a distorting effect on the ranking. Asia's startup ecosystems are a prominent example of this. In these places, ecosystems have been established based on a governance model different from the US pattern (*Klingler-Vidra* [2018]), or (particularly in the case of China) a specific model of state-university-industry cooperation within the ecosystem has been developed (*Lyu et al.* [2019], *Yang–Welch* [2012], *Yu et al.* [2021], *Zhuang et al.* [2021]). In this operational logic, for example, the degree of internet censorship and entrepreneurial freedom are not determinants of ecosystem development but rather of the dominant exit strategy and the industry focus of firms. A good illustration of this is the Chinese practice where startup exits due to acquisitions by large state-owned enterprises or university holdings (such as Tsinghua Holding) is much more common than exits due to IPOs. An important difference between the methodologies of the two rankings is that StartupBlink considers the number of local branches/offices of international corporations among the indicators of ecosystem quality. This is an objective indicator that is not included in the Startup Genome indicators, yet the presence of large international companies has been proven to be a boosting factor for innovation ecosystems (*Etzkowitz* [2002]).

---

[8] https://www.startupblink.com/startups

---

[9] The DBI in its current form will not be published in the future (see https://www.reuters.com/business/world-bank-taking-steps-boost-research-integrity-after-data-rigging-scandal-2021-10-11/).

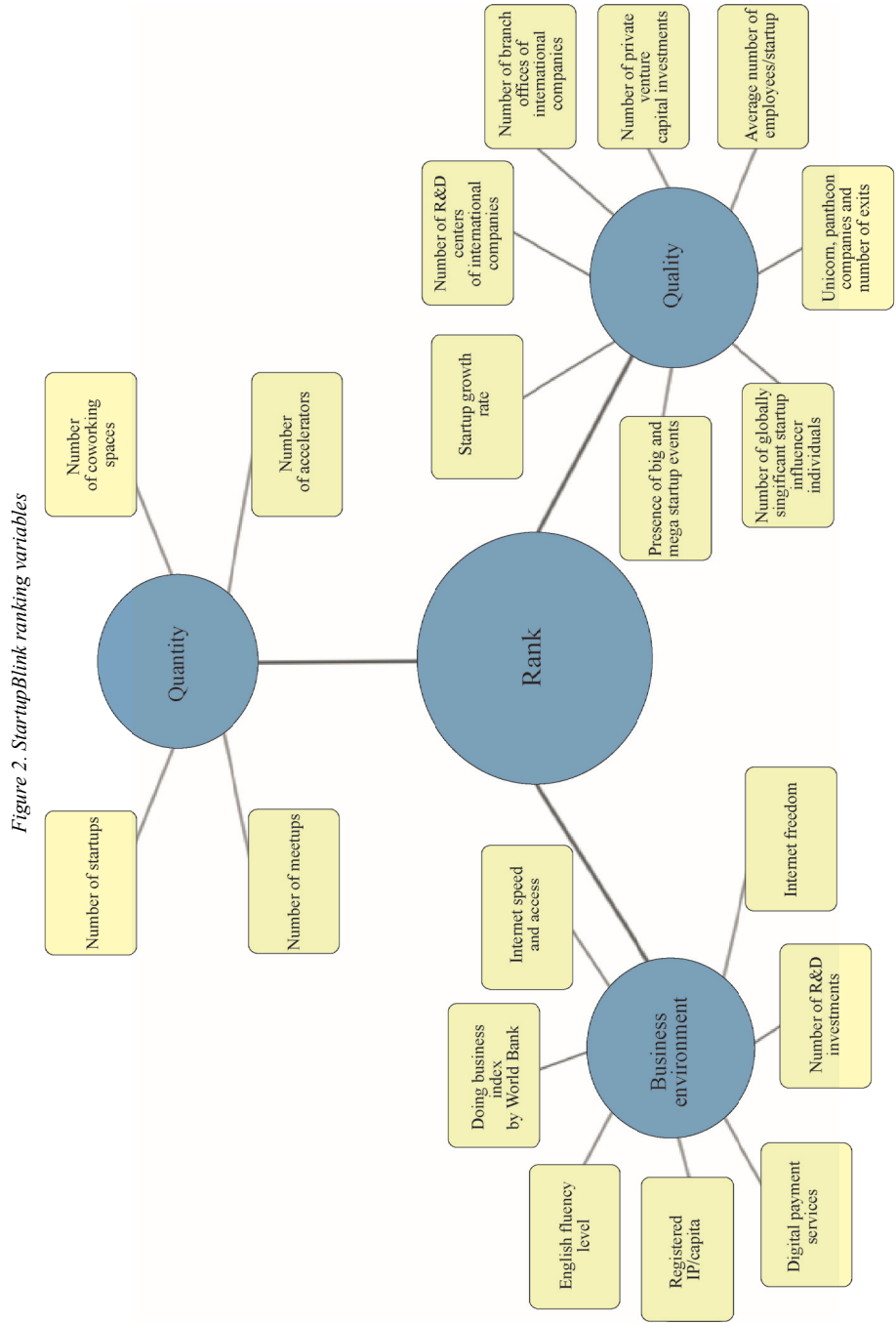

*Figure 2. StartupBlink ranking variables*

## 4. Comparison of two startup ecosystem rankings

### 4.1. Digital performance of the rankings

The strong digital presence of startup companies and actors in the startup ecosystem as well as the digital nature of the industry, is a well-known phenomenon. It is precisely the digital technology focus (in addition to operating in an extremely uncertain business environment) that distinguishes startups from other early-stage companies. As startup ecosystem rankings provide information for this specific target group, it might be worth including a comparison of the web performance of each ranking in the analysis, which is becoming an increasingly important impact indicator as digitalisation becomes more prevalent. Analyses of this type – without claiming to be exhaustive – have already been conducted for higher education institutions (*Lee–Park* [2012]), university libraries (*Vallez–Ventura* [2020]), and economic operators (*Wang–Vaughan* [2014]).

In this respect, an important measure is to examine the popularity of the websites publishing the given ecosystem ranking. To filter out 'digital noise' (e.g. automatic exchange of data between sites, the activity of search engines and other robots), it is not enough to rely on clickstream data alone. For this reason, one of the most accepted measures of site popularity, in addition to Google Search PageRank, is Amazon's Alexa rank[10], which measures the number of visitors to a site and the engagement of visitors with the site and can filter for machine activity. Alexa rank is a ratio scale measure of the popularity of web pages, derived from the website visit data of Alexa toolbar users and the engagement rate (the percentage of return visits and activity) for that page in 120 days from the date of ranking. Alexa only scans the main page of a given site; it is not suitable for analysing the traffic of individual subpages. Examining the Alexa ranks of the StartupBlink and Startup Genome sites, we can see that the popularity of the StartupBlink site is several times higher than that of the Startup Genome site. (See Figure 3.) However, this may also be because StartupBlink functions as both a ranking and user platform.

The number of links to each site (see Figure 4) and the starting point of traffic to each site (see Figure 5) are also important aspects to examine. In this respect, Startup Genome performs better as it has more links to this site and the starting points of its traffic are more diversified than those of StartupBlink.

---

[10] https://blog.alexa.com/marketing-research/alexa-rank/

*Figure 3. Alexa rank derived from traffic data for the websites of the rankings analysed*

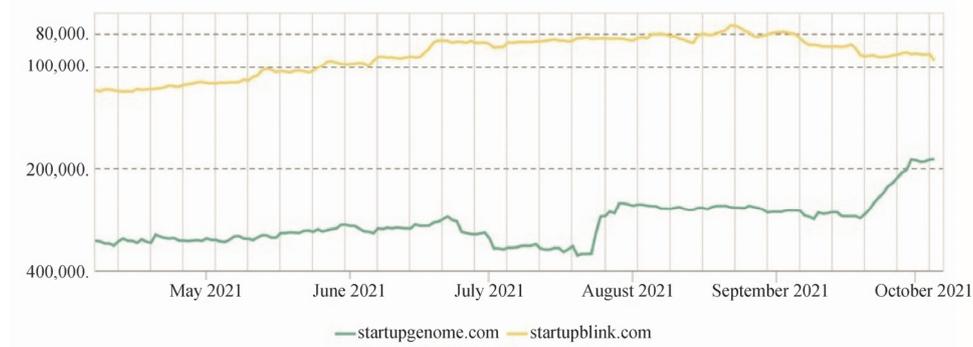

*Source*: Own query using Amazon's Alexa.

*Figure 4. Number of external links to the websites of the rankings analysed*

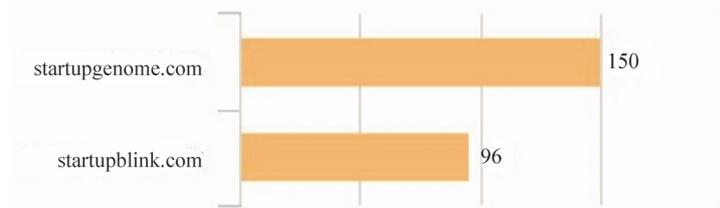

*Source*: Own query using Amazon's Alexa.

*Figure 5. Sources of traffic for the rankings analysed*

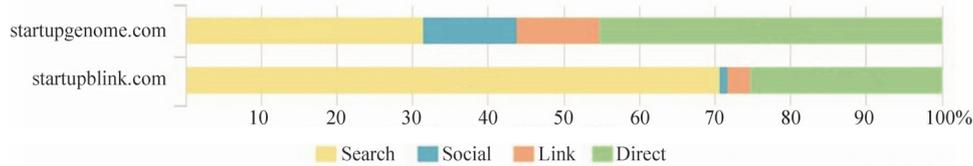

*Source*: Own query using Amazon's Alexa.

## 4.2. Comparison of the rankings

When comparing the two rankings, we face the problem that one consists of 140 elements (Startup Genome) and the other of 1,000 elements (StartupBlink). On this basis, we can confidently adopt the hypothesis that the 140-element ranking will completely overlap with the 1,000-element ranking. Following a brief analysis, this will be confirmed. After grammatically combining the names of the ecosystems (e.g. Silicon Valley = San Francisco Bay Area, Research Triangle = Raleigh, Greater Helsinki = Helsinki), there is only one item in the 2020 Startup Genome ranking that is missing from the StartupBlink list. This is specifically the ecosystem around the Chinese city of Wuxi. The reason for this shortfall is that StartupBlink considers the cities of Wuxi and Changzhou, the latter being already on the list as one ecosystem. More useful correlations can be obtained by including only the first 140 items of the StartupBlink list in the analysis. In this case, the results presented in Table 3 are derived for the overlap.

Table 3

*Degree of overlap for the top 140 places in the two rankings*

| Denomination | Number of overlaps |
| --- | --- |
| Top 10 | |
| Startup Genome | 10 |
| StartupBlink | 10 |
| Top 50 | |
| Startup Genome | 49 |
| StartupBlink | 49 |
| Top 100 | |
| Startup Genome | 87 |
| StartupBlink | 91 |
| Complete | |
| Startup Genome | 110 |
| StartupBlink | 110 |

The results show that there are 110 overlapping ecosystems on the 140-element list of the two rankings, 30 ecosystems that are not on the Startup Genome 140-element list and 30 ecosystems that are on the Startup Genome 140-element list but ranked higher than 140 on the StartupBlink ranking. It is noteworthy that the proportion of overlapping items decreases as you move backwards in the rankings, with more than half of the non-overlapping items being in the range between the 100$^{th}$ and 140$^{th}$ positions. The decreasing overlap clearly shows a deterioration in the reliability of the rankings, which becomes more significant from rank 90 onwards. The reason for this is to be found in the variables on which the rankings are based, as presented earlier, and the associated weights. Since the number of startups, investments, and successful exits do not follow a normal distribution (i.e. a significant proportion of them are concentrated in ecosystems at the top of the rankings), the variables that have a fundamental impact on the ranking (e.g. number of successful exits above $50 million) gradually take on the value of 0 for ecosystems at the bottom of the rankings, giving way to variables that normally have low weight (e.g. number of accelerators, internet freedom, DBI, GEDI[11] index, GDP/capita). Accordingly, the ranking methodology (as it cannot handle an increasing number of 0 values for the variables) also becomes less efficient and eventually invalid/meaningless. A similar (though not nearly as drastic) overlap pattern can be observed in the higher education rankings (*Aguillo et al.* [2010] p. 249). In practical terms, this means that a startup ecosystem actor can draw almost no meaningful conclusions from the fact that its host ecosystem is ranked 120th or 138th in a given ranking. Thus, an important and pragmatically relevant feature of the startup ecosystem rankings is that their meaningful information content is gradually and after the 100th position drastically reduced for users.

The analysis of the correlation of the rankings is particularly relevant in this case, given that the methodology of one of the rankings under analysis (StartupBlink) is only partially transparent due to the lack of information on the weights of the variables. The correlations in question can only be examined in terms of the elements covered by the two rankings ($N = 110$), so only these elements from the complete sample were used in the calculation.

Table 4

*Correlations between the overlapping elements of the 140-element sample in the two rankings*

| Kendall's tau-b | StartupBlink/Startup Genome | Correlation coefficient | 0.611** |
| | | Significance (2-tailed) | 0.000 |
| | | $N$ | 110 |
| Spearman's rho | StartupBlink/Startup Genome | Correlation coefficient | 0.796** |
| | | Significance (2-tailed) | 0.000 |
| | | $N$ | 110 |

** Correlation is significant at the 0.01 level (2-tailed).

The results reveal strong positive correlations that are, however, reduced by outlier-type values, which are explained by different weighting and indicators. (See Figure 6.) Nevertheless, in this case, it may be worthwhile to investigate the trends of the differences by time series analysis in the future.

---

[11] GEDI: global ecosystem dynamics investigation.

*Figure 6. Standard deviation of the overlapping elements of the 140-element sample in the two rankings*

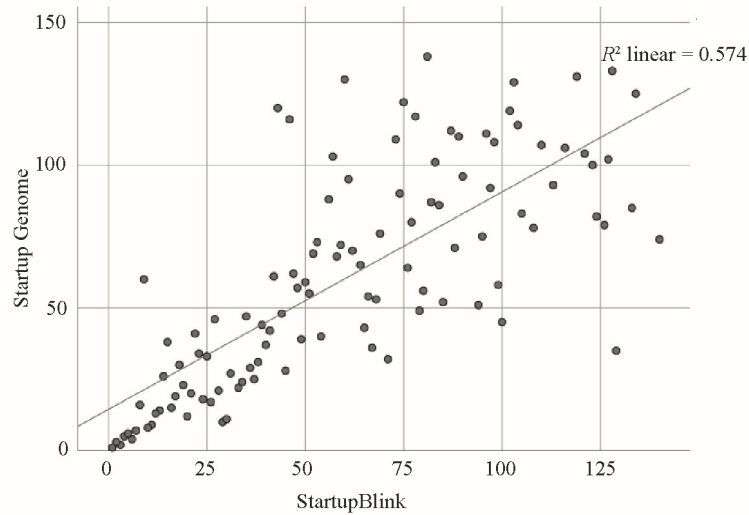

*Source*: Own elaboration using IBM SPSS (v.26).

In the case of the top 100 elements of the two rankings, we can witness a slight reduction in the outliers (see Figure 7) accompanied by strengthening of the correlation (see Table 5). Here, the 9th place (StartupBlink) and 60th place (Startup Genome) of the Moscow startup ecosystem need to be explained based on a more detailed analysis. The reasons for the large difference also require a time series analysis as it is almost impossible to identify the causes of the significant deviation based on data for a single year, which could be induced by 1-2 high-value internal startup exit events (e.g. management buy-out) not 'detected' by either of the two rankings. In the case of outliers, it is worth observing the trends, which allow us to determine with a high degree of certainty whether they are caused by a one-off bias or whether (in the case of trend-like differences) we are witnessing the so-called 'gaming' effect or the operation of Campbell's law[12]. There are many examples of this phenomenon in other rankings, perhaps the best known among them is the comet-like rise of Georgia in the World Bank's DBI ranking (*Cooley–Snyder* [2015] pp. 151–177).

---

[12] The stronger the impact of a social indicator (or ranking) in social decision-making is, the more exposed it is to corruption pressures and, at the same time, the less suitable it becomes for measuring the social phenomena it was originally designed to assess.

Table 5

*Correlations between the overlapping elements of the 100-element sample in the two rankings*

| Kendall's tau-b | StartupBlink/Startup Genome | Correlation coefficient | 0.641** |
|---|---|---|---|
| | | Significance (2-tailed) | 0.000 |
| | | N | 78 |
| Spearman's rho | StartupBlink/Startup Genome | Correlation coefficient | 0.819** |
| | | Significance (2-tailed) | 0.000 |
| | | N | 78 |

\*\* The correlation is significant at the 0.01 level (2-tailed).

*Figure 7. Standard deviation of the overlapping elements of the 100-element sample in the two rankings*

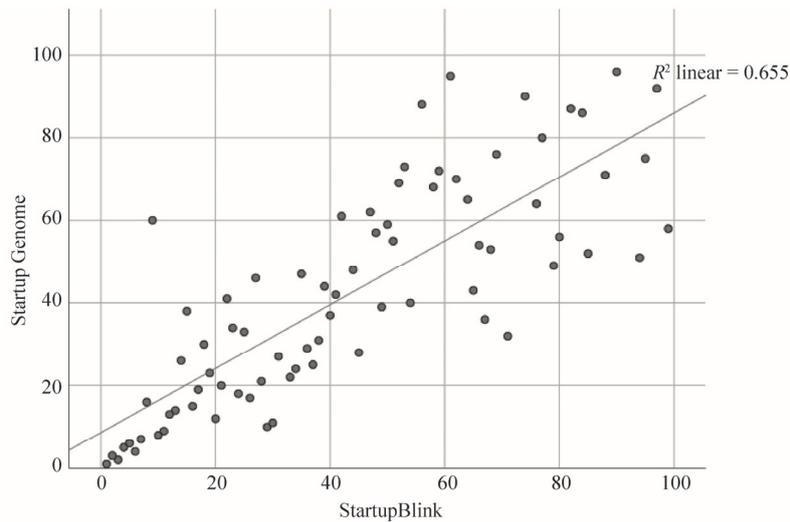

*Source*: Own elaboration using IBM SPSS (v.26).

## 5. Summary of results, future research directions

The present analysis undertook to address three research questions for which the results are detailed in the previous sections. The findings related to each research question are as follows:

*1*. There are currently several regional, sectoral rankings on the web that are closely related to the startup theme but only two global rankings of startup ecosystems can be identified, Startup Genome and StartupBlink. These correspond to the criteria previously outlined for the rankings, that is, they are public, electronically accessible, and searchable by the target group.

*2*. The two rankings are typically based on data available in digital data sources as well as on other rankings and indices and the composite variables derived from them. In terms of methodology, the rankings are challenging to compare (due to the limited methodological transparency of one of them). The variables used in the rankings are predominantly variables defined by the resource-focused approach to ecosystem concepts (quantitative and qualitative); variables and indicators on ecosystem interactions and governance structures are rarely found. For this reason, the latter characteristics of ecosystems can only be inferred indirectly from the rankings examined.

*3*. A strong correlation between the identified startup ecosystem rankings can be observed, with a decreasing trend towards the lower tracts. The overlap of ecosystems in the rankings is significant, with the degree of overlap declining in the second half of each ranking. Despite the strong correlation, there are several cases where a substantial difference in the positions of individual ecosystems in the rankings can be observed.

Although the analysis has provided answers to the defined research questions, these answers raise many additional issues that require further investigation on the subject and have far-reaching policy relevance. Further research is required to obtain scientific knowledge that can be measured against other (higher education, innovation, economic) rankings of startup ecosystems.

Time series analyses constitute an adequate tool for examining trends in ecosystem rankings. This requires a multi-year study, which may explain the specific outliers and their patterns among the rankings observed in the present study. The comparison is possible from 2017 as data for both rankings are available from this year onwards. Thus, a 5-year time span can now be analysed, which, based on experience, is already sufficient to identify different trends.

This research did not undertake a deep statistical analysis between the rankings due to space constraints; however, this must be done in the future to improve understanding. In this respect, it is particularly important to accurately determine the distance between the positions of ecosystems in each ranking, for which adequate methodology is already available.

Given that a considerable part of variables used by the rankings relies on different electronic databases (e.g. Crunchbase, Orb Intelligence, PitchBook), the quantity and quality of the data they contain have a fundamental impact on the ranking of individual ecosystems. It is, therefore, an essential line of research to define the extent to which the use of these databases is widespread in the various ecosystems and whether their use is supported or even mandated by public policy (which induces more data to be generated in these databases, giving a better ranking to the particular ecosystem).

One of the fundamental reasons for compiling the rankings is to support and underpin decisions. For this reason, a relevant issue for startup ecosystem rankings is the extent to which they influence the decisions of individual actors and contribute to, for example, the choice of location for individual startup companies or the decisions of investors in the venture capital world. Further qualitative studies are needed in the future to determine this.

In conclusion, we can state that in the few years since their launch the startup ecosystem rankings have gained a significant degree of publicity on the World Wide Web. This is demonstrated by the traffic figures of the websites displaying these rankings. Nevertheless, scientific interest has not yet been developed in this regard. In the case of other rankings (see higher education), however, the first related scientific research started several years after the first rankings had been published. If the interest shown by ecosystem stakeholders in these rankings persists, more and more research will likely focus on exploring their characteristics and potential impacts. This is also justified because these rankings (and the underlying methodology and variables) can contribute to a better understanding and assessment of the functioning of entrepreneurial ecosystems in a broader sense. Despite the startup focus (i.e. the fact that these rankings consider the perspective of startups and their investors), the rankings and the underlying analyses/databases assess highly relevant elements of the innovation potential of a given region. For this reason, startup ecosystem rankings can provide valuable inputs for future research to explore the functioning and efficiency of entrepreneurial ecosystems in micro- and macro-geographical terms and the entrepreneurial discovery process in a particular region.